\documentclass{article}
\usepackage{amsfonts}
\usepackage{amssymb}
\begin{document}
\begin{flushleft}
A Criterion for Unentanglement of Quantum States\\
\vskip 5pt
Xufeng Liu\\
School of Mathematical Sciences, Peking University\\
Beijing 100871, P.R.China\\
e-mail: liuxf@pku.edu.cn\\
\end{flushleft}

\noindent Abstract. The concept of entanglement is at the core of the theory of quantum information. In this paper a criterion for unentanglement of quantum states is proposed and proved. This criterion is natural, practical and easy to check.
\vskip 5pt
\noindent Mathematics Subject Classification (2020): 46N50.
\vskip 5pt
\noindent Key words: quantum state, unentanglement
\vskip 10pt
\noindent 1. Preliminaries
\vskip 5pt
This section mainly serves to fix the notation.

Let $\Lambda$ be a set. The set consisting of all the finite subsets of $\Lambda$ is denoted by ${\cal F}(\Lambda)$. For all $A, B\in {\cal F}(\Lambda)$ we define $A\ge B$ to mean $B\subset A$, then the relation $\ge $ is transitive and reflexive on ${\cal F}(\Lambda)$ and enjoys the property: for $A$ in ${\cal F}(\Lambda)$ and $B$ in ${\cal F}(\Lambda)$ there is $C$ in ${\cal F}(\Lambda)$ such that $C\ge A$ and $C\ge B$. In other words, the relation $\ge $ directs the set ${\cal F}(\Lambda)$.

Let ${\cal X}$ be a Banach space and $\{x_{\alpha},\alpha\in\Lambda\}$ a family in ${\cal X}$. If the net $\{\sum_{\alpha\in A}x_{\alpha}, A\in {\cal F}(\Lambda)\}$ converges to $x$ in ${\cal X}$, we say this family to be summable and call $
x$ its sum and denote the sum by $\sum_{\alpha\in\Lambda}x_{\alpha}$.
Throughout this paper inner products of different Hilbert spaces will be denoted by the same $\langle,\rangle$ and the convention is adopted that the inner product associated with a complex Hilbert space is complex linear in the second argument and conjugate linear in the first argument. Let ${\cal H}$ be a (complex or real) Hilbert space. We denote the sets of all finite-rank linear operators, all bounded linear operators and all compact linear operators on ${\cal H}$ by ${\cal F}({\cal H})$, ${\cal L}({\cal H})$ and ${\cal C}({\cal H})$ respectively and we use ${\cal S(H)}$, ${\cal T(H)}$ and ${\cal L_{\rm s}(H)}$ to stand for the set of all Hilbert-Schmidt operators, the set of all trace class operators and the set of all bounded self-adjoint operators on ${\cal H}$ respectively.
A bounded self-adjoint operator A on ${\cal H}$ is called non-negative if $\langle A\phi,\phi\rangle\ge 0$ for all $\phi\in {\cal H}$, where $\langle,\rangle$ is the inner product on ${\cal H}$.  The set of all non-negative bounded self-adjoint operators on ${\cal H}$ is denoted by ${\cal L_{\rm s}(H)}^{+}$. For $A\in {\cal L(H)}$ we put $|A|=(A^*A)^{1/2}$.

Let ${\cal H}_1, {\cal H}_2,\cdots, {\cal H}_n$ be Hilbert spaces and $\odot_{k=1}^{n}{\cal H}_k={\cal H}_1\odot{\cal H}_2\odot\cdots\odot{\cal H}_n$ their algebraic tensor product. Then the formula
$$
\bigg<\sum_{i=1}^{m}\alpha_i(\otimes_{k=1}^{n}\phi_{ki}),\sum_{j=1}^{l}\beta_j(\otimes_{k=1}^{n}\varphi_{kj})\bigg>=\sum_{i,j}\bar{\alpha_i}\beta_j\prod\nolimits_{k=1}^{n}\big\langle\phi_{ki},\varphi_{kj}\big\rangle,
$$
where $\alpha_i,\beta_j\in{\mathbb C}$ and $\phi_{ki},\varphi_{kj}\in {\cal H}_k$, defines an inner product on $\odot_{k=1}^{n}{\cal H}_k$. The completion of $\odot_{k=1}^{n}{\cal H}_k$ for the associated norm is called the tensor product of ${\cal H}_1, {\cal H}_2,\cdots, {\cal H}_n$ and denoted by ${\cal H}_1\otimes{\cal H}_2\otimes\cdots\otimes{\cal H}_n$ or $\otimes_{k=1}^{n}{\cal H}_k$.

Now let us present some well known conclusions for later use.

Suppose that $\{e_{i\alpha_i},\alpha_i\in\Lambda_i\}(i=1,2,\cdots,n)$ are complete orthonormal sets in the Hilbert spaces ${\cal H}_i(i=1,2,\cdots,n)$ respectively, then $\{\otimes_{i=1}^{n}e_{i\alpha_i},\alpha_i\in\Lambda_i,i=1,2,\cdots,n\}$ is a complete orthonormal set in $\otimes_{i=1}^{n}{\cal H}_i$. If $A_i\in {\cal L}({\cal H}_i)(i=1,2,\cdots,n)$ there is a unique operator in $\otimes_{k=1}^{n}{\cal H}_k$, which is denoted by $A_1\otimes A_2\otimes\cdots\otimes A_n$ or $\otimes_{k=1}^{n}A_k$, such that
$$
(\otimes_{k=1}^{n}A_k)(\otimes_{k=1}^{n}\phi_{k})=\otimes_{k=1}^{n}(A_k\phi_{k})
$$
for all $\phi_{k}\in {\cal H}_k(k=1,2,\cdots,n)$ and we have $\|\otimes_{k=1}^{n}A_k\|=\prod_{k=1}^{n}\|A_k\|$.

For an arbitrary Hilbert space ${\cal H}$  the set ${\cal C}({\cal H})$ is a linear space enjoying, among others, the following properties: ${\cal F}({\cal H})$ is dense in ${\cal C}({\cal H})$ and ${\cal C}({\cal H})$ closed in ${\cal L}({\cal H})$ relative to the operator norm topology; if $A,B\in {\cal L}({\cal H})$  then $AB$ lies in ${\cal C}({\cal H})$ when $A$ or $B$ is compact.

Let $A$ be an operator in ${\cal C(H)}$. Then $|A|$ is also in ${\cal C(H)}$. Let $\{\lambda_n, n\in\Lambda\}$ be the at most countable set of eigenvalues of $|A|$, repeated according to their multiplicity (necessarily finite). By definition, the operator $A$ belongs to ${\cal T(H)}$ if and only if 
$$
\|A\|_1=\sum_{n\in\Lambda}\lambda_n<\infty,
$$ 
and belongs to ${\cal S(H)}$ if and only if 
$$
\|A\|_2^2=\sum_{n\in\Lambda}\lambda_n^2<\infty.
$$
Operators in ${\cal T(H)}$ and in ${\cal S(H)}$ are called trace class and Hilbert-Schmidt operators respectively. 

The set ${\cal T(H)}$ is a linear space and the above defined $\|\cdot\|_1$ is a norm, called trace norm, on ${\cal T(H)}$. With respect to the trace norm ${\cal T(H)}$ is a Banach space and ${\cal F}({\cal H})$ is dense in ${\cal T(H)}$. If $A$ lies in ${\cal T(H)}$ then the sum 
$$
{\rm tr }[A]=\sum_{\alpha\in\Lambda}\langle Ae_{\alpha},e_{\alpha}\rangle
$$
is independent of the chosen orthonormal basis (complete orthonormal set) $\{e_{\alpha},\alpha\in\Lambda\}$ in ${\cal H}$. Moreover, we have $\|A\|\le \|A\|_1$ for all $A\in {\cal T(H)}$ and 
$$
\|AB\|_1\le \|A\|_1\|B\|,\ \|BA\|_1\le \|B\|\cdot\|A\|_1
$$
for all $A\in {\cal T(H)}$ and $B\in {\cal L(H)}$. If $A\in {\cal T}({\cal H}_1)$ and $B\in {\cal T}({\cal H}_2)$ then $A\otimes B\in{\cal T}({\cal H}_1\otimes{\cal H}_2)$, $\|A\otimes B\|_1=\|A\|_1\|B\|_1$ and ${\rm tr  }[A\otimes B]={\rm tr  }[A]{\rm tr  }[B]$.

The set ${\cal S(H)}$ is a Hilbert space with the inner product $\langle,\rangle$ (by abuse of notation) defined by the formula
$$
\langle A,B\rangle=\sum_{\alpha\in\Lambda}\langle Ae_{\alpha},Be_{\alpha}\rangle
$$
for all $A,B\in {\cal S(H)}$. Here $\{e_{\alpha},\alpha\in\Lambda\}$ is an arbitrary orthonormal basis in ${\cal H}$. This inner product is compatible with the above defined $\|\cdot\|_2$ in the sense that $\|A\|_2^2=\langle A,A\rangle$ for all $A\in {\cal S(H)}$.
We have $\|A\|_2\le\|A\|_1$ for all $A\in {\cal C(H)}$ and thus ${\cal T(H)}\subset{\cal S(H)}$. Moreover, if $A,B\in {\cal S(H)}$ then $AB^*\in {\cal T(H)}$ and $\langle A,B\rangle={\rm tr  }[AB^*]$.

\vskip 10pt
\noindent 2. Schmidt's Decomposition 
\vskip 5pt
To the best of my knowledge, Schmidt's decomposition has its origin in [1] and its finite dimensional version has become well known since the rise of quantum information. Despite this fact a complete proof of this decomposition will be presented here in the general framework of tensor product of not necessarily separable Hilbert spaces, for the sake of methodology needed in proving the main results of this paper. The following procedure is based on the mathematical discussions on this topic carried out in [2] in the context of quantum measurement and in the framework of concrete Hilbert spaces.  

Definition 2.1. Let ${\cal H}_1$ and ${\cal H}_2$ be Hilbert spaces. A conjugate linear map $A:{\cal H}_1\rightarrow {\cal H}_2$ is defined to be a map such that 
$$
A(\alpha_1\phi_1+\alpha_2\phi_2)=\bar{\alpha}_1A(\phi_1)+\bar{\alpha_2}A(\phi_2)
$$
for all $\alpha_1,\alpha_2\in {\mathbb C}$ and $\phi_1,\phi_2\in {\cal H}_1$ and the set of all bounded conjugate linear maps from ${\cal H}_1$ to ${\cal H}_2$ is denoted by $\bar{\cal L}({\cal H}_1,{\cal H}_2)$.

For a $\Phi\in {\cal H}_1\otimes {\cal H}_2$ and all $\varphi_1\in {\cal H}_1,\varphi_2\in {\cal H}_2$ we have
$$|\langle \Phi,\varphi_1\otimes \varphi_2\rangle|\le \|\Phi\|\cdot\|\varphi_1\otimes \varphi_2\|= \|\Phi\|\cdot\|\varphi_1\|\cdot\|\varphi_2\|.
$$ 
This fact justifies the following definition.

Definition 2.2. Let $\Phi$ be a vector in ${\cal H}_1\otimes {\cal H}_2$. We define $T_{\Phi}\in \bar{\cal L}({\cal H}_1,{\cal H}_2)$ and $T^{\Phi}\in \bar{\cal L}({\cal H}_2,{\cal H}_1)$ such that
$$
\langle T_{\Phi}(\varphi_1),\varphi_2\rangle=\langle \Phi,\varphi_1\otimes \varphi_2\rangle=\langle T^{\Phi}(\varphi_2),\varphi_1\rangle
$$
for all $\varphi_1\in {\cal H}_1$ and $\varphi_2\in {\cal H}_2$. 

Proposition 2.1. For all $\varphi_1\in {\cal H}_1, \varphi_2\in {\cal H}_2$ and all $\Phi\in {\cal H}_1\otimes {\cal H}_2$ we have $\|T_{\Phi}(\varphi_1)\|\le \|\Phi\|\cdot\|\varphi_1\|$ and $\|T^{\Phi}(\varphi_2)\|\le \|\Phi\|\cdot\|\varphi_2\|$.

Proof. By definition 
$$\|T_{\Phi}(\varphi_1)\|^2=|\langle \Phi,\varphi_1\otimes T_{\Phi}(\varphi_1)\rangle|\le \|\Phi\|\cdot\|\varphi_1\|\cdot\|T_{\Phi}(\varphi_1)\|,$$
it then follows directly that $\|T_{\Phi}(\varphi_1)\|\le \|\Phi\|\cdot\|\varphi_1\|$. The second conclusion can be proved in the same way.

For $\Phi\in{\cal H}_1\otimes {\cal H}_2$, by definition the operators $T_{\Phi}$ and $T^{\Phi}$ lie in $\bar{\cal L}({\cal H}_1,{\cal H}_2)$ and $\bar{\cal L}({\cal H}_2,{\cal H}_1)$ respectively. As a result, the operator $T^{\Phi}T_{\Phi}$ lies in ${\cal L}({\cal H}_1)$.

Proposition 2.2. We have 
$$
\|T^{\Phi}T_{\Phi}-T^{\Psi}T_{\Psi}\|\le (\|\Phi\|+\|\Psi\|)\|\Phi-\Psi\|
$$
for $\Phi,\Psi\in{\cal H}_1\otimes {\cal H}_2$.

Proof. In fact, for all $\phi,\varphi\in{\cal H}_1$ we have
\begin{eqnarray*}
|\langle (T^{\Phi}T_{\Phi}-T^{\Psi}T_{\Psi})\phi,\varphi\rangle|&=&|\langle \Phi,\varphi\otimes T_{\Phi}\phi\rangle-\langle \Psi,\varphi\otimes T_{\Psi}\phi\rangle|\\
&\le&|\langle \Phi-\Psi,\varphi\otimes T_{\Phi}\phi\rangle|+|\langle \Psi,\varphi\otimes(T_{\Phi}-T_{\Psi})\phi\rangle|\\
&\le&(\|\Phi\|+\|\Psi\|)\|\Phi-\Psi\|\cdot\|\phi\|\cdot\|\varphi\|,
\end{eqnarray*}
the proposition thus follows.

Proposition 2.3. Let $\Phi$ be a vector in ${\cal H}_1\otimes {\cal H}_2$. If there exist finite orthogonal sets $\{\phi_{\alpha},\alpha\in\Lambda_1\}\subset {\cal H}_1$ and $\{\varphi_{\beta},\beta\in\Lambda_2\}\subset {\cal H}_2$ such that $\Phi$ lies in the linear subspace of ${\cal H}_1\otimes {\cal H}_2$ spanned $\{\phi_{\alpha}\otimes\varphi_{\beta},\alpha\in\Lambda_1,\beta\in\Lambda_2\}$ the operator $T^{\Phi}T_{\Phi}$ is of finite rank.

Proof. Denote the smallest linear subspace of ${\cal H}_2$ containing $\{\varphi_{\beta},\beta\in\Lambda_2\}$ by $W_{\Lambda_2}$ and its orthogonal complement in ${\cal H}_2$ by $W_{\Lambda_2}^{\perp}$. Then we have 
$$
\langle T_{\Phi}\phi,\varphi\rangle=\langle\Phi,\phi\otimes\varphi\rangle=0
$$
for all $\phi\in {\cal H}_1$ and $\varphi\in W_{\Lambda_2}^{\perp}$, and thus 
$$
T_{\Phi}({\cal H}_1)\subset(W_{\Lambda_2}^{\perp})^{\perp}=W_{\Lambda_2}.
$$
This means $\dim(T_{\Phi}({\cal H}_1))<\infty$, and the proposition follows immediately.

Proposition 2.4. The operators $T^{\Phi}T_{\Phi}$ and $T_{\Phi}T^{\Phi}$ belongs to ${\cal C}({\cal H}_1)\cap{\cal L}_{\rm s}({\cal H}_1)^{+}$ and ${\cal C}({\cal H}_2)\cap{\cal L}_{\rm s}({\cal H}_2)^{+}$ respectively for all $\Phi$ in ${\cal H}_1\otimes {\cal H}_2$.

Proof. For $\Phi\in{\cal H}_1\otimes {\cal H}_2$, there exist orthogonal sets $\{\phi_{\alpha},\alpha\in\Delta_1\}\subset {\cal H}_1$ and $\{\varphi_{\beta},\beta\in\Delta_2\}\subset {\cal H}_2$ such that 
$$
\Phi=\sum_{\alpha\in\Delta_1\atop\beta\in\Delta_2}\phi_{\alpha}\otimes\varphi_{\beta},
$$
and for all $\epsilon>0$, there exist finite subsets $\Lambda_1\subset\Delta_1$ and $\Lambda_2\subset\Delta_2$ such that 
$$
\|\Phi-\sum_{\alpha\in\Lambda_1\atop\beta\in\Lambda_2}\phi_{\alpha}\otimes\varphi_{\beta}\|<\epsilon.
$$
It then follows from Propositions 2.2 and 2.3 that the operator $T^{\Phi}T_{\Phi}$ can be approximated by finite rank operators in the operator norm topology. Consequently we have $T^{\Phi}T_{\Phi}\in {\cal C}({\cal H}_1)$ thanks to the fact that ${\cal F}({\cal H}_1)\subset{\cal C}({\cal H}_1)=\overline{{\cal C}({\cal H}_1)}$. Here $\overline{{\cal C}({\cal H}_1)}$ means the operator norm closure of ${\cal C}({\cal H}_1)$. That $T^{\Phi}T_{\Phi}$ lies in ${\cal L}_{\rm s}({\cal H}_1)^{+}$ is obvious:
\begin{eqnarray*}
&&\langle T^{\Phi}T_{\Phi}(\phi),\varphi\rangle=\langle \Phi,\varphi\otimes T_{\Phi}(\phi)\rangle=\langle T_{\Phi}(\varphi),T_{\Phi}(\phi)\rangle=\overline{\langle T_{\Phi}(\phi),T_{\Phi}(\varphi)\rangle},\\
&&\langle \phi,T^{\Phi}T_{\Phi}(\varphi)\rangle=\overline{\langle T^{\Phi}T_{\Phi}(\varphi),\phi\rangle}=\overline{\langle \Phi,\phi\otimes T_{\Phi}(\varphi)\rangle}=\overline{\langle T_{\Phi}(\phi),T_{\Phi}(\varphi)\rangle},\\
&&\langle T^{\Phi}T_{\Phi}(\phi),\phi\rangle=\overline{\langle T_{\Phi}(\phi),T_{\Phi}(\phi)\rangle}=\|T_{\Phi}(\phi)\|^2\ge 0
\end{eqnarray*}
for all $\phi,\varphi\in{\cal H}_1$. The proof of the first conclusion is thus completed. The second conclusion can be proved in the same way.

Lemma 2.1. Let $\Phi$ be a vector in ${\cal H}_1\otimes {\cal H}_2$ and $\{e_{\alpha},\alpha\in\Lambda\}$ an orthonormal basis of ${\cal H}_1$. We have the representation
$$
\Phi=\sum_{\alpha\in\Lambda}e_{\alpha}\otimes T_{\Phi}(e_{\alpha})
$$
if the family $\{e_{\alpha}\otimes T_{\Phi}(e_{\alpha}),\alpha\in\Lambda\}$ is summable.

Proof. In fact, if the family $\{e_{\alpha}\otimes T_{\Phi}(e_{\alpha})\}$ is summable, then for an arbitrary orthonormal basis $\{f_{\beta},\beta\in\Delta\}$ in ${\cal H}_2$ we have 
\begin{eqnarray*}
\bigg<\sum_{\alpha\in\Lambda}e_{\alpha}\otimes T_{\Phi}(e_{\alpha}),e_{\gamma}\otimes f_{\beta}\bigg>&=&\sum_{\alpha\in\Lambda}\langle e_{\alpha},e_{\gamma}\rangle\langle T_{\Phi}(e_{\alpha}),f_{\beta}\rangle\cr
&=&\langle T_{\Phi}(e_{\gamma}),f_{\beta}\rangle=\langle {\Phi},e_{\gamma}\otimes f_{\beta}\rangle
\end{eqnarray*}
for all $\gamma\in\Lambda$ and $\beta\in\Delta$. This proves the lemma.

Proposition 2.5. Let $\Phi$ be a vector in ${\cal H}_1\otimes {\cal H}_2$. Then there exist at most countable orthonormal sets $\{e_{\alpha},\alpha\in\Lambda\}\subset {\cal H}_1$, $\{f_{\alpha},\alpha\in\Lambda\}\subset {\cal H}_2$ and set of positive numbers $\{\lambda_{\alpha},\alpha\in\Lambda\}$ such that
$$
\Phi=\sum_{\alpha\in\Lambda}\sqrt{\lambda_{\alpha}}\,e_{\alpha}\otimes f_{\alpha},\ \  \sum_{\alpha\in\Lambda}\lambda_{\alpha}=\|\Phi\|^2.
$$

Proof. We have $T^{\Phi}T_{\Phi}$ and $T_{\Phi}T^{\Phi}\in {\cal C}({\cal H}_1)\cap{\cal L}_{\rm s}({\cal H}_1)^{+}$. It then follows that there exist an orthonormal basis $\{e_{\alpha},\alpha\in\Lambda\}\cup\{e_{\beta},\beta\in\Delta\}$ of ${\cal H}_1$ and positive numbers $\lambda_{\alpha}(\alpha\in\Lambda)$ such that
$$
T^{\Phi}T_{\Phi}(e_{\alpha})=\lambda_{\alpha}e_{\alpha},\  T^{\Phi}T_{\Phi}(e_{\beta})=0
$$
for all $\alpha\in\Lambda$ and $\beta\in\Delta$. Here $\Lambda$ is an at most countable set. As an immediate result, we have $\|T_{\Phi}e_{\alpha}\|^2=\lambda_{\alpha}>0$ and $T_{\Phi}e_{\beta}=0$ for all $\alpha\in\Lambda, \beta\in\Delta$. Moreover, $\{T_{\Phi}(e_{\alpha})/\|T_{\Phi}(e_{\alpha})\|,\alpha\in\Lambda\}$ is an orthonormal set in ${\cal H}_2$. Let $f_{\alpha}=T_{\Phi}(e_{\alpha})/\|T_{\Phi}(e_{\alpha})\|(\alpha\in\Lambda)$. Then we obtain
$$
\|\Phi\|^2\ge\sum_{\alpha\in\Lambda}\|\langle \Phi,e_{\alpha}\otimes f_{\alpha}\rangle\|^2=\sum_{\alpha\in\Lambda}\lambda_{\alpha}=\sum_{\alpha\in\Lambda}\|e_{\alpha}\otimes T_{\Phi}(e_{\alpha})\|^2.
$$
This proves the summability of $\{e_{\alpha}\otimes T_{\Phi}(e_{\alpha}),\alpha\in\Lambda\cup\Delta\}$, and with it the proposition thanks to Lemma 2.1.

The representation of $\Phi$ proved in Proposition 2.5 is usually referred to as Schmidt's decomposition. The Schmidt's decomposition of $\Phi$ is unique in the sense that the involved $\lambda_{\alpha}'$s, $e_{\alpha}'$s and $f_{\alpha}'$s must obey the  conditions: $T^{\Phi}T_{\Phi}(e_{\alpha})=\lambda_{\alpha}e_{\alpha}, T_{\Phi}T^{\Phi}(f_{\alpha})=\lambda_{\alpha}f_{\alpha}$. 

Now let us conclude this section by a proposition, which can be regarded as a slight generalization of Schmidt's decomposition. 

Proposition 2.6. Assume that $\Phi\in {\cal H}_1\otimes {\cal H}_2$ satisfies the condition: there exist complete orthonormal sets $\{e_{\alpha},\alpha\in\Delta_1\}$ and $\{f_{\beta},\beta\in\Delta_2\}$ in ${\cal H}_1$ and ${\cal H}_2$ respectively such that the family $\{|\langle \Phi,e_{\alpha}\otimes f_{\beta}\rangle|,\alpha\in\Delta_1,\beta\in\Delta_2\}$ is summable. Then
$$
\Phi=\sum_{\nu\in\Lambda}e_{\nu}\otimes T_{\Phi}(e_{\nu})
$$
for an arbitrary orthonormal basis $\{e_{\nu},\nu\in\Lambda\}$ in ${\cal H}_1$.

Proof. Thanks to Lemma 2.1, we only have to prove the summability of the family $\{e_{\nu}\otimes T_{\Phi}(e_{\nu}),\alpha\in\Lambda\}$. The proof proceeds as follows. The assumption of the proposition guarantees the existence of orthogonal sets $\{\phi_{\alpha},\alpha\in\Delta_1\}\subset {\cal H}_1$ and $\{\varphi_{\beta},\beta\in\Delta_2\}\subset {\cal H}_2$ such that 
$$
\Phi=\sum_{\alpha\in\Delta_1\atop\beta\in\Delta_2}\phi_{\alpha}\otimes\varphi_{\beta},\ \ \sum_{\alpha\in\Delta_1\atop\beta\in\Delta_2}\|\phi_{\alpha}\|\cdot\|\varphi_{\beta}\|<\infty.
$$ 
Thus we have
\begin{eqnarray*}
\|e_{\nu}\otimes T_{\Phi}(e_{\nu})\|^2&=&\langle T_{\Phi}(e_{\nu}),T_{\Phi}(e_{\nu})\rangle=|\langle {\Phi},e_{\nu}\otimes T_{\Phi}(e_{\nu})\rangle|\\
&=&\bigg|\sum_{\gamma,\beta}\langle \phi_{\gamma},e_{\nu}\rangle\langle \varphi_{\beta},T_{\Phi}(e_{\nu})\rangle\bigg|\\
&\le&\sum_{\gamma,\beta}|\langle \phi_{\gamma},e_{\nu}\rangle|\cdot|\langle \varphi_{\beta},T_{\Phi}(e_{\nu})\rangle|
\end{eqnarray*}
for all $\nu\in\Lambda$. Then for any finite subset $\Lambda_1$ of $\Lambda$
\begin{eqnarray*}
&&\sum_{\nu\in\Lambda_1}\|e_{\nu}\otimes T_{\Phi}(e_{\nu})\|^2\le\sum_{\gamma,\beta}\sum_{\nu\in\Lambda_1}|\langle \phi_{\gamma},e_{\nu}\rangle|\cdot|\langle \varphi_{\beta},T_{\Phi}(e_{\nu})\rangle|\\
&&\le\sum_{\gamma,\beta}\bigg(\sum_{\nu\in\Lambda_1}|\langle \phi_{\gamma},e_{\nu}\rangle|^2\bigg)^{1/2}\bigg(\sum_{\nu\in\Lambda_1}|\langle \varphi_{\beta},T_{\Phi}(e_{\nu})\rangle|^2\bigg)^{1/2}\\
&&=\sum_{\gamma,\beta}\bigg(\sum_{\nu\in\Lambda_1}|\langle \phi_{\gamma},e_{\nu}\rangle|^2\bigg)^{1/2}\bigg(\sum_{\nu\in\Lambda_1}|\langle T^{\Phi}(\varphi_{\beta}),e_{\nu}\rangle|^2\bigg)^{1/2}\\
&&\le\sum_{\gamma,\beta}\|\phi_{\gamma}\|\cdot\|T^{\Phi}(\varphi_{\beta})\|\le\|\Phi\|\sum_{\gamma,\beta}\|\phi_{\gamma}\|\cdot\|\varphi_{\beta}\|.
\end{eqnarray*}
So the family $\{\|e_{\nu}\otimes T_{\Phi}(e_{\nu})\|^2,\nu\in\Lambda\}$ is summable. This implies the summability of the family $\{e_{\nu}\otimes T_{\Phi}(e_{\nu}),\nu\in\Lambda\}$ since it is an orthogonal set in ${\cal H}_1\otimes {\cal H}_2$ and hence completes the proof.

\vskip 10pt
\noindent 3. Criterion for Product Pure State
\vskip 5pt
Definition 3.1. Let ${\cal H}$ be a Hilbert space. A vector $\Phi$ in ${\cal H}$ is called a pure state of ${\cal H}$ if $\|\Phi\|=1$. Let ${\cal H}={\cal H}_1\otimes {\cal H}_2$. A pure state $\Phi$ of ${\cal H}$ is called a product pure state if there exist $\varphi_1\in {\cal H}_1$ and $\varphi_2\in {\cal H}_2$ such that $\Phi=\varphi_1\otimes \varphi_2$. 

Definition 3.2. Let ${\cal H}$ be a Hilbert space. An operator $P\in {\cal L_{\rm s}(H)}$ is called a projector upon $P{\cal H}$ if $P^2=P$. Let $\phi$ be a nonzero vector in ${\cal H}$. Denote by $\hat{\phi}$ the unit vector $\phi/\|\phi\|$. The operator $P_{\phi}\in{\cal L(H)}$ is defined to be the projector upon the one dimensional linear subspace of ${\cal H}$ spanned by $\phi$, and called a rank $1$ projector. The totalities of rank $1$ projectors and of projectors on ${\cal H}$ are denoted by ${\cal P}_1({\cal H})$ and ${\cal P}({\cal H})$ respectively.

Proposition 3.1. For nonzero vectors $\phi,\varphi\in {\cal H}$ we have $\|P_{\phi}-P_{\varphi}\|\le (\|\hat{\phi}\|+\|\hat{\varphi}\|)\|\hat{\phi}-\hat{\varphi}\|$.

Proof. By definition
$$
(P_{\phi}-P_{\varphi})\psi=\langle \hat{\phi},\psi\rangle\hat{\phi}-\langle \hat{\varphi},\psi\rangle\hat{\varphi}
$$
for all $\psi\in {\cal H}$. Hence we have
\begin{eqnarray*}
\|(P_{\phi}-P_{\varphi})\psi\|&\le&\|(\hat{\phi}-\hat{\varphi})\langle \hat{\phi},\psi\rangle\|+\|\langle \hat{\phi}-\hat{\varphi},\psi\rangle\hat{\varphi}\|\\
&\le&(\|\hat{\phi}\|+\|\hat{\varphi}\|)\|\hat{\phi}-\hat{\varphi}\|\cdot\|\psi\|
\end{eqnarray*} 
for all $\psi\in {\cal H}$. The proposition then follows directly.

Definition 3.3. Let ${\cal H}$ be a Hilbert space. We use the symbol ${\mathbf 1}$ to stand for the identity operator on ${\cal H}$.

Proposition 3.2. A unit vector $\Phi$ in a Hilbert space ${\cal H}_1\otimes {\cal H}_2$ is a product pure state of ${\cal H}_1\otimes {\cal H}_2$ if
$$
\langle \Phi,(P\otimes Q)\Phi\rangle=\langle \Phi,(P\otimes {\mathbf 1})\Phi\rangle\langle \Phi,({\mathbf 1}\otimes Q)\Phi\rangle
$$
for all $P\in {\cal P}_1({\cal H}_1),Q\in {\cal P}_1({\cal H}_2)$.  

Proof. Take the Schmidt's decomposition 
$$
\Phi=\sum_{\alpha\in\Lambda}\sqrt{\lambda_{\alpha}}\,e_{\alpha}\otimes f_{\alpha}
$$
and apply the condition stated in the proposition to the projectors $P=P_{e_{\lower 2pt\hbox{$\scriptscriptstyle\beta$}}}$ and $Q=P_{f_{\beta}}$. We then come straightforwardly to the conclusion that $\lambda_{\beta}^2=\lambda_{\beta}$ for all $\beta\in\Lambda$, namely, $\lambda_{\beta}$ can only take the value $0$ or $1$ for all $\beta\in\Lambda$. But we have $\sum_{\alpha\in\Lambda}\lambda_{\alpha}=\|\Phi\|^2=1$, so $\Lambda$ must be a single point set. The proposition is thus proved.

Lemma 3.1. Let $\Phi$ be a vector in the Hilbert space ${\cal H}=\otimes_{k=1}^{n}{\cal H}_k$ and ${\mathbf h}_k\subset {\cal H}_k (k=1,2,\cdots,n)$ finite dimensional subspaces. If $\Phi$ lies in $\otimes_{k=1}^{n}{\mathbf h}_k$, then $P_{\Phi}$ has the representation
$$
P_{\Phi}=\sum_{i_1\cdots i_n}\alpha_{i_1\cdots i_n}( P_{i_1}\otimes\cdots\otimes P_{i_n})
$$
where $\alpha_{i_1\cdots i_n}\in {\mathbb C}$, $P_{i_j}\in {\cal P}_1({\cal H}_j)(j=1,2,\cdots,n)$ and the summation indices run over some finite sets.

Proof. Let ${\mathbf h}=\otimes_{k=1}^{n}{\mathbf h}_k$. Then we have $P_{\Phi}({\mathbf h}^{\perp})\subset{\mathbb C}\Phi\subset{\mathbf h}$ by definition and $P_{\Phi}({\mathbf h}^{\perp})\subset{\mathbf h}^{\perp}$ by the self-adjointness of $P_{\Phi}$. This leads to $P_{\Phi}({\mathbf h}^{\perp})=\{0\}$. 

On the other hand, we notice that $P_{\Phi}\!\left |_{\mathbf h}\right.$, the restriction of $P_{\Phi}$ to ${\mathbf h}$, lies in ${\cal L({\mathbf h})}$, so it can be expressed as a linear combination of operators of the form $A_{j_1}\otimes\cdots\otimes A_{j_n}$, where $A_{j_i}\in {\cal L}({\mathbf h}_i)(i=1,2,\cdots,n)$. But every operator in ${\cal L}({\mathbf h}_i)$, which can be written as a linear combination of two self-adjoint operators on ${\mathbf h}_i$, can be represented as a linear combination of operators from ${\cal P}_1({\mathbf h}_i)$, so we conclude that $P_{\Phi}\!\left |_{\mathbf h}\right.$ can be represented as follows:
$$
P_{\Phi}\!\left |_{\mathbf h}\right.=\sum_{i_1\cdots i_n}\alpha_{i_1\cdots i_n}( \raise 2pt\hbox{$p$}_{i_1}\otimes\cdots\otimes \raise 2pt\hbox{$p$}_{i_n})
$$
where $\raise 2pt\hbox{$p$}_{i_j}\in {\cal P}_1({\mathbf h}_j)(j=1,2,\cdots,n)$.

Now define $P_{i_j}\in {\cal L}({\cal H}_j)(j=1,2,\cdots,n)$ by
$$
P_{i_j}\big|_{{\mathbf h}_j}=\raise 2pt\hbox{$p$}_{i_j},\ P_{i_j}\big|_{{\mathbf h}_j^{\perp}}=0.
$$
It is easy to check that $P_{i_j}\in {\cal P}_1({\cal H}_j)(j=1,2,\cdots,n)$. We claim that
$$
P_{\Phi}=\sum_{i_1\cdots i_n}\alpha_{i_1\cdots i_n}( P_{i_1}\otimes\cdots\otimes P_{i_n}).
$$ 
Indeed, by definition we already have
$$
P_{\Phi}\!\left|_{\mathbf h}\right.=\sum_{i_1\cdots i_n}\alpha_{i_1\cdots i_n}( P_{i_1}\otimes\cdots\otimes P_{i_n})\!\left|_{\mathbf h}\right..
$$ 
It remains to show that the two operators are also identical on ${\mathbf h}^{\perp}$. In fact one can readily check that $(P_{i_1}\otimes\cdots\otimes P_{i_n})({\mathbf h}^{\perp})\subset{\mathbf h}\cap{\mathbf h}^{\perp}$. Hence the two operators are both zero operators on ${\mathbf h}^{\perp}$. The claim thus follows and the proof is completed.

Definition 3.4. Let ${\cal H}_1,{\cal H}_2,\cdots{\cal H}_n(n\ge 2)$ be Hilbert spaces. For $A_i\in{\cal L}({\cal H}_i)(i=1,2,\cdots,n)$ we define 
$$\bar{A}_i\in{\cal L}(\otimes_{k=1}^{n}{\cal H}_k)(i\ge 1),\ \ \widetilde{A}_i\in{\cal L}(\otimes_{k=2}^{n}{\cal H}_k)(i\ge 2)
$$ 
by the formulae:
\begin{eqnarray*}
&&\bar{A}_i={\mathbf 1}\otimes\cdots\otimes{\mathbf 1}\otimes A_i\otimes {\mathbf 1}\otimes\cdots\otimes{\mathbf 1}\  (\hbox{$A_i$ appearing at the $i$'th position}),\\
&&\widetilde{A}_i={\mathbf 1}\otimes\cdots\otimes{\mathbf 1}\otimes A_i\otimes {\mathbf 1}\otimes\cdots\otimes{\mathbf 1}\  (\hbox{$A_i$ appearing at the $(i-1)$'th position}).
\end{eqnarray*}

It may be desirable to illustrate this definition by a simple example. Take $n=4$ , $\phi_i\in {\cal H}_i(i=1,2,3,4)$ and $A_2\in{\cal L}({\cal H}_2)$. Then by the above definition
\begin{eqnarray*}
&&\bar{A}_2(\phi_1\otimes\phi_2\otimes\phi_3\otimes\phi_4)=\phi_1\otimes (A_2\phi_2)\otimes\phi_3\otimes\phi_4,\\
&&\widetilde{A}_2(\phi_2\otimes\phi_3\otimes\phi_4)=(A_2\phi_2)\otimes\phi_3\otimes\phi_4.
\end{eqnarray*}

Lemma 3.2. Let ${\cal H}_1,{\cal H}_2,\cdots,{\cal H}_n(n\ge 2)$ be Hilbert spaces and $\Phi$ a unit vector in $\otimes_{k=1}^{n}{\cal H}_k$. If
$$
\prod_{i=1}^{n}\langle \Phi,\overline{P}_i\Phi\rangle=\langle \Phi,(P_1\otimes P_2\otimes\cdots\otimes P_n)\Phi\rangle
$$
for all $P_i\in {\cal P}_1({\cal H}_i)(i=1,2,\cdots,n)$, then
$$
\prod_{i=2}^{n}\langle \Phi,({\mathbf 1}\otimes\widetilde{P}_i)\Phi\rangle=\langle \Phi,({\mathbf 1}\otimes P_2\otimes\cdots\otimes P_n)\Phi\rangle
$$
for all $P_i\in {\cal P}_1({\cal H}_i)(i=2,\cdots,n)$.

Proof. Let $\{e_{\alpha},\alpha\in\Lambda\}$ be an orthonormal basis of ${\cal H}_1$.  Taking $P_1=P_{e_{\alpha}}$, by definition and from the condition of the lemma we obtain 
\begin{eqnarray*}
&&\langle \Phi,(P_{e_{\alpha}}\otimes{\mathbf 1}\otimes\cdots\otimes{\mathbf 1})\Phi\rangle\prod_{i=2}^{n}\langle \Phi,({\mathbf 1}\otimes\widetilde{P}_i)\Phi\rangle\cr
&&=\langle \Phi,(P_{e_{\alpha}}\otimes P_2\otimes\cdots\otimes P_n)\Phi\rangle
\end{eqnarray*}
for all $\alpha\in\Lambda,P_i\in {\cal P}_1({\cal H}_i)(i=2,\cdots,n)$. Since ${\mathbf 1}=\sum_{\alpha\in\Lambda}P_{e_{\alpha}}$ in the strong topology on ${\cal H}_1$, it then follows that
\begin{eqnarray*}
\prod_{i=2}^{n}\langle \Phi,({\mathbf 1}\otimes\widetilde{P}_i)\Phi\rangle&=&\langle \Phi,({\mathbf 1}\otimes{\mathbf 1}\otimes\cdots\otimes{\mathbf 1})\Phi\rangle\prod_{i=2}^{n}\langle \Phi,({\mathbf 1}\otimes\widetilde{P}_i)\Phi\rangle\cr
&=&\langle \Phi,(\sum_{\alpha}P_{e_{\alpha}}\otimes{\mathbf 1}\otimes\cdots\otimes{\mathbf 1})\Phi\rangle\prod_{i=2}^{n}\langle \Phi,({\mathbf 1}\otimes\widetilde{P}_i)\Phi\rangle\cr
&=&\sum_{\alpha}\langle \Phi,(P_{e_{\alpha}}\otimes{\mathbf 1}\otimes\cdots\otimes{\mathbf 1})\Phi\rangle\prod_{i=2}^{n}\langle \Phi,({\mathbf 1}\otimes\widetilde{P}_i)\Phi\rangle\cr
&=&\sum_{\alpha}\langle \Phi,(P_{e_{\alpha}}\otimes P_2\otimes\cdots\otimes P_n)\Phi\rangle\cr
&=&\langle \Phi,(\sum_{\alpha}P_{e_{\alpha}}\otimes P_2\otimes\cdots\otimes P_n)\Phi\rangle\cr
&=&\langle \Phi,({\mathbf 1}\otimes P_2\otimes\cdots\otimes P_n)\Phi\rangle,
\end{eqnarray*}
and the lemma is proved.

Definition 3.5. Let $\Phi$ be a vector in the Hilbert space $\otimes_{k=1}^{n}{\cal H}_k(n\ge 2)$. If
$$
\prod_{i=1}^{n}\langle \Phi,\overline{P}_i\Phi\rangle=\langle \Phi,(P_1\otimes P_2\otimes\cdots\otimes P_n)\Phi\rangle
$$
for all $P_i\in {\cal P}_1({\cal H}_i)(i=1,2,\cdots,n)$, we say that $\Phi$ satisfies the unentanglement condition of $\otimes_{k=1}^{n}{\cal H}_k$.

Lemma 3.3. Let $\Phi$ be a unit vector in the Hilbert space $\otimes_{k=1}^{n}{\cal H}_k(n\ge 3)$. If $\Phi$ satisfies the unentanglement condition of $\otimes_{k=1}^{n}{\cal H}_k$, then viewed as a vector in the Hilbert space ${\cal H}_1\otimes ({\cal H}_2\otimes\cdots\otimes {\cal H}_n)$ it also satisfies the unentanglement condition of this latter Hilbert space. 

Proof. What we need to prove is 
$$
\langle \Phi,(P_1\otimes{\mathbf 1})\Phi\rangle\langle \Phi,({\mathbf 1}\otimes P)\Phi\rangle=\langle \Phi,(P_1\otimes P)\Phi\rangle
$$
for all $P_1\in {\cal P}_1({\cal H}_1),P\in {\cal P}_1(\otimes_{i=2}^n{\cal H}_i)$. The proof of this condition proceeds in steps.

(1) The operator $P$ assumes the form $\otimes_{i=2}^{n}P_i,P_i\in {\cal P}_1({\cal H}_i)(i=2,\cdots,n)$. According to Lemma 3.2, we have
$$
\langle \Phi,({\mathbf 1}\otimes P)\Phi\rangle=\prod_{i=2}^{n}\langle \Phi,({\mathbf 1}\otimes\widetilde{P}_i)\Phi\rangle
$$
and thus
\begin{eqnarray*}
&&\langle \Phi,(P_1\otimes{\mathbf 1})\Phi\rangle\langle \Phi,({\mathbf 1}\otimes P)\Phi\rangle=\prod_{i=1}^{n}\langle \Phi,\overline{P}_i\Phi\rangle\cr
&&=\langle \Phi,(P_1\otimes P_2\otimes\cdots\otimes P_n)\Phi\rangle=\langle \Phi,(P_1\otimes P)\Phi\rangle
\end{eqnarray*}
by the hypothesis of the lemma.

(2) The operator $P$ is a linear combination of operators of the form $\otimes_{i=2}^{n}P_i,P_i\in {\cal P}_1({\cal H}_i)(i=2,\cdots,n)$. In this case, the condition is also true, thanks to its linearity with respect to the argument $P$ and the conclusion of (1).

(3) The operator $P$ takes the form $P_{\Psi},\Psi\in\otimes_{k=2}^{n}{\cal H}_k$. This is the general case. Let $\{e_{i\alpha_i},\alpha_i\in\Lambda_i\}\subset{\cal H}_i(i=2,\cdots,n)$ be orthonormal bases. Then
$$
\Psi=\sum_{\alpha_2}\cdots\sum_{\alpha_n}\langle \otimes_{i=2}^{n}e_{i\alpha_i},\Psi\rangle(\otimes_{i=2}^{n}e_{i\alpha_i}).
$$
For finite sets $\Delta_i\subset\Lambda_i(i=2,\cdots,n)$ we define
$$
\Psi(\Delta_2,\cdots,\Delta_n)=\sum_{\alpha_2\in\Delta_2}\cdots\sum_{\alpha_n\in\Delta_n}\langle \otimes_{i=2}^{n}e_{i\alpha_i},\Psi\rangle(\otimes_{i=2}^{n}e_{i\alpha_i}).
$$
and denote by ${\mathbf h}_i$ the subspace of ${\cal H}_i$ spanned by $\{e_{i\alpha_i},\alpha_i\in\Delta_i\}$ for $i\in \{2,\cdots,n\}$. By definition $\Psi(\Delta_2,\cdots,\Delta_n)$ lies in ${\mathbf h}_2\otimes\cdots\otimes{\mathbf h}_n$ and it is clear that $\Psi$ can be approximated by such vectors. Then it follows from Lemma 3.1 and Proposition 3.1 that the operator $P$ can be approximated by operators of the form considered in (2) and this allows us to obtain what we need by the standard technique of taking limit.

Theorem 3.1.  A unit vector $\Phi$ in the Hilbert space $\otimes_{k=1}^{n}{\cal H}_k(n\ge 2)$ is a product pure state if and only if it satisfies the unentanglement condition of $\otimes_{k=1}^{n}{\cal H}_k$.

Proof. The ``only if" part is trivial. We prove the ``if part" by induction on $n$. When $n=2$, the conclusion is true thanks to Proposition 3.2. Suppose that the conclusion is true for $n=\ell$. Let $\Phi$ be a unit vector satisfying the unentanglement condition of $\otimes_{k=1}^{\ell+1}{\cal H}_k$. Then according to Lemma 3.3, it satisfies the unentanglement condition of ${\cal H}_1\otimes ({\cal H}_2\otimes\cdots\otimes {\cal H}_{\ell+1})$. So Proposition 3.2 works again and insures the existence of unit vectors $\phi\in {\cal H}_1$ and $\Psi\in {\cal H}_2\otimes\cdots\otimes {\cal H}_{\ell+1}$ such that $\Phi=\phi\otimes\Psi$. Now by virtue of Lemma 3.2 one can easily check that $\Psi$ satisfies the unentanglement condition of ${\cal H}_2\otimes\cdots\otimes {\cal H}_{\ell+1}$. Consequently, by the induction hypothesis $\Psi$ is a product pure state of ${\cal H}_2\otimes\cdots\otimes {\cal H}_{\ell+1}$. This completes the proof. 

\vskip 10pt
\noindent 4. Criterion for Product Mixed State
\vskip 5pt
Let ${\cal H}_1$ and ${\cal H}_2$ be Hilbert spaces. Then ${\cal S}({\cal H}_1)$
and ${\cal S}({\cal H}_2)$ are Hilbert spaces (not necessarily separable even if ${\cal H}_1$ and ${\cal H}_2$ are separable).

Proposition 4.1. Let $R$ be an operator in ${\cal S}({\cal H}_1)\otimes {\cal S}({\cal H}_2)$. Then there exist at most countable orthonormal sets $\{A_{\alpha},\alpha\in\Lambda\}\subset {\cal S}({\cal H}_1)$, $\{B_{\alpha},\alpha\in\Lambda\}\subset {\cal S}({\cal H}_2)$ and set of positive numbers $\{\lambda_{\alpha},\alpha\in\Lambda\}$ such that
$$
R=\sum_{\alpha\in\Lambda}\sqrt{\lambda_{\alpha}}\,A_{\alpha}\otimes B_{\alpha},\ \  \sum_{\alpha\in\Lambda}\lambda_{\alpha}=\|R\|_2^2.
$$
If $R$ is self-adjoint then all $A_{\alpha},B_{\alpha}$ can be chosen to be self-adjoint.

Proof. The application of Proposition 2.5 to ${\cal S}({\cal H}_1)\otimes {\cal S}({\cal H}_2)$ directly yields the first conclusion. For the second conclusion, notice that ${\cal S}({\cal H}_1)\cap{\cal L}_s({\cal H}_1)$ and ${\cal S}({\cal H}_2)\cap{\cal L}_s({\cal H}_2)$ are real Hilbert spaces and that the same proposition can be applied to their tensor product, which is also a real Hilbert space. The conclusion then follows immediately.

Definition 4.1. Let ${\cal H}_i(i=1,\cdots,n)$ be Hilbert spaces. An operator $T\in {\cal S}({\cal H}_1)\otimes\cdots\otimes {\cal S}({\cal H}_n)$ is called a generic trace class operator on ${\cal H}_1\otimes\cdots\otimes{\cal H}_n$ if there exist $\{T_{i\alpha_i},\alpha_i\in\Lambda_i\}\subset {\cal T}({\cal H}_i)(i=1,\cdots,n)$ such that 
$$
T=\sum_{\alpha_1}\cdots\sum_{\alpha_n}T_{1\alpha_1}\otimes\cdots\otimes T_{n\alpha_n}
$$
where $\Lambda_i(i=1,\cdots,n)$ are finite sets.

Proposition 4.2. Let $R$ be a generic trace class operator on ${\cal H}_1\otimes {\cal H}_2$. Then there exist finite sets $\{A_1,\cdots,A_r\}\subset {\cal T}({\cal H}_1)$, $\{B_1,\cdots,B_r\}\subset {\cal T}({\cal H}_2)$ and set of positive numbers $\{\lambda_1,\cdots,\lambda_r\}$ such that 
$$
{\rm tr}[A_iA_j^*]={\rm tr}[B_iB_j^*]=\delta_{ij},\ i,j=1,\cdots,r
$$
and 
$$
R=\sum_{i=1}^{r}\sqrt{\lambda_{i}}\,A_{i}\otimes B_{i},\ \  \sum_{i=1}^{r}\lambda_{i}=\|R\|_2^2={\rm tr}[RR^*].
$$

Proof. Notice that ${\rm tr}[A_iA_j^*]$  is none other than $\langle A_i,A_j\rangle$ and likewise, ${\rm tr}[B_iB_j^*]=\langle B_i,B_j\rangle$. It remains to show that $R$ being a generic trace class operator on ${\cal H}_1\otimes {\cal H}_2$ Proposition 4.1 has the following stronger conclusions: $\Lambda$ is a finite set and $A_{\alpha}\in {\cal T}({\cal H}_1), B_{\alpha}\in {\cal T}({\cal H}_2)$ for all $\alpha\in\Lambda$. In fact, when $R$ is a generic trace class operator on ${\cal H}_1\otimes {\cal H}_2$ it assumes the form $R=\sum_{i=1}^{m}S_i\otimes T_i$, where $S_i\in {\cal T}({\cal H}_1),T_i\in{\cal T}({\cal H}_2)(i=1,\cdots,m)$. It follows that the operator $T^RT_R$ is of finite rank and this implies the first conclusion, namely the finiteness of the set $\Lambda$, which is defined in Proposition 4.1. For the second conclusion, just observe that
\begin{eqnarray*}
&&\|T_R(A_{\alpha})\|B_{\alpha}=T_R(A_{\alpha})=\sum_{i=1}^{m}\langle A_{\alpha},S_i\rangle T_i,\\
&&\|T_R(B_{\alpha})\|A_{\alpha}=T^R(B_{\alpha})=\sum_{i=1}^{m}\langle B_{\alpha},T_i\rangle S_i
\end{eqnarray*}
where $A_{\alpha}$ and $B_{\alpha}$ keep the same meanings as in Proposition 4.1. Namely, the operators $A_{\alpha}$ and $B_{\alpha}$ are both linear combinations of trace class operators. Then the conclusion follows at once.

The decomposition of $R$ presented in Proposition 4.2 will be referred to as Schmidt's decomposition of~$R$.

Definition 4.2. Let ${\cal H}$ be a Hilbert space. An operator $\rho\in {\cal T}({\cal H})\cap {\cal L_{\rm s}(H)}^{+}$ is called a mixed state of ${\cal H}$ if ${\rm tr  }[\rho]=1$.

Definition 4.3. A generic mixed state $\rho$ of ${\cal H}_1\otimes\cdots\otimes{\cal H}_n$ is defined to be a mixed state that can be written as a finite linear combination of operators of the form $\rho_1\otimes\cdots\otimes\rho_n$, where $\rho_i$ is a mixed state of ${\cal H}_i\ (i=1,\cdots,n)$.

Notice that for a mixed state $\rho$ we have $\|\rho\|_1={\rm tr  }[\rho]=1$ and a generic mixed state is necessarily a generic trace class operator.

Definition 4.4. A mixed state $\rho$ of ${\cal H}_1\otimes\cdots\otimes{\cal H}_n$ is called a product mixed state if it is of the form $\rho_1\otimes\cdots\otimes\rho_n$, $\rho_i$ being a mixed state of ${\cal H}_i\ (i=1,\cdots,n)$.

Lemma 4.1. Let $\rho$ be a trace class operator on ${\cal H}_1\otimes {\cal H}_2$. If
$$
{\rm tr  }[(P\otimes Q)\rho]={\rm tr  }[(P\otimes {\mathbf 1})\rho]{\rm tr  }[({\mathbf 1}\otimes Q)\rho]
$$
for all $P\in {\cal P}_1({H}_1),Q\in {\cal P}_1({H}_2)$ then it is still true for all $P\in {\cal C}({\cal H}_1),Q\in {\cal C}({\cal H}_2)$.

Proof. First, it is easily seen that when the operators $P$ and $Q$ can be written as linear combinations of rank $1$ projectors we can derive the equation 
$$
{\rm tr  }[(P\otimes Q)\rho]={\rm tr  }[(P\otimes {\mathbf 1})\rho]{\rm tr  }[({\mathbf 1}\otimes Q)\rho]
$$
from the hypothesis. 

Next, let $Q$ be a linear combination of rank $1$ projectors and $P$ a compact operator. Then for all $\epsilon>0$ there exists a $P^{\prime}\in {\cal L}({\cal H}_1)$, which is a linear combination of rank $1$ projectors, such that $\|P-P^{\prime}\|<\epsilon$ and we have
\begin{eqnarray*}
&&\bigl|{\rm tr  }[(P\otimes Q)\rho]-{\rm tr  }[(P\otimes {\mathbf 1})\rho]{\rm tr  }[({\mathbf 1}\otimes Q)\rho]\bigr|\\
&&=\big|{\rm tr  }[(P\otimes Q)\rho]-{\rm tr  }[(P^{\prime}\otimes Q)\rho]+{\rm tr }[(P^{\prime}\otimes Q)\rho]-{\rm tr }[(P\otimes {\mathbf 1})\rho]{\rm tr }[({\mathbf 1}\otimes Q)\rho]\big|\\
&&\le \big|{\rm tr }[((P-P^{\prime})\otimes Q)\rho]\big|+\big|{\rm tr }[(P^{\prime}\otimes Q)\rho]-{\rm tr }[(P\otimes {\mathbf 1})\rho]{\rm tr }[({\mathbf 1}\otimes Q)\rho]\big|\\
&&=\big|{\rm tr }[((P-P^{\prime})\otimes Q)\rho]\big|+\big|{\rm tr }[((P^{\prime}-P)\otimes {\mathbf 1})\rho]{\rm tr }[({\mathbf 1}\otimes Q)\rho]\big|\\
&&\le\|P-P^{\prime}\|\cdot\|Q\|\cdot\|\rho\|_1+\|P^{\prime}-P\|\cdot\|\rho\|_1\|Q\|\cdot\|\rho\|_1\cr
&&=\|P-P^{\prime}\|\cdot\|Q\|(\|\rho\|_1+\|\rho\|_1^2)\\
&&<\epsilon\|Q\|(\|\rho\|_1+\|\rho\|_1^2)=2\epsilon\|Q\|.
\end{eqnarray*}
This means that for a fixed, but arbitrary $Q$ as defined above (linear combination of rank $1$ projectors), the lemma is true for all $P\in {\cal C}({\cal H}_1)$.

Finally, let $P\in {\cal C}({\cal H}_1)$ and $Q\in {\cal C}({\cal H}_2)$ and for a fixed, but arbitrary $P$ apply the preceding argument to $Q$. Then, in view of the just established conclusion, the lemma follows directly. 

Proposition 4.3. Let $\rho$ be a generic mixed state of ${\cal H}_1\otimes {\cal H}_2$. If
$$
{\rm tr }[(P\otimes Q)\rho]={\rm tr }[(P\otimes {\mathbf 1})\rho]{\rm tr }[({\mathbf 1}\otimes Q)\rho]
$$
for all $P\in {\cal P}_1({\cal H}_1),Q\in {\cal P}_1({\cal H}_2)$ then it is a product mixed state.

Proof. Take Schmidt's decomposition of $\rho$:
$$
\rho=\sum_{i=1}^{r}\sqrt{\lambda_{i}}\,A_{i}\otimes B_{i},
$$
and recall that here $A_i\in{\cal T}({\cal H}_1),B_i\in{\cal T}({\cal H}_2)$ and 
$$
{\rm tr }[A_iA_j^*]={\rm tr }[B_iB_j^*]=\delta_{ij}
$$
 for all $i,j\in\{1,\cdots,r\}$. Lemma 4.1 allows us to take $P=A_i^*$ and $Q=B_i^*$ in the condition of this proposition. Now we have, on the one hand, 
$$
{\rm tr }[(A_i^*\otimes B_i^*)\rho]={\rm tr }[(A_i^*\otimes {\mathbf 1})\rho]{\rm tr }[({\mathbf 1}\otimes B_i^*)\rho],
$$
and on the other hand,
\begin{eqnarray*}
&&{\rm tr }[(A_i^*\otimes {\mathbf 1})\rho]=\sum_{k=1}^r\sqrt{\lambda_{k}}{\rm tr }[A_i^*A_k]{\rm tr }[B_k]=\sqrt{\lambda_{i}}{\rm tr }[B_i],\\
&&{\rm tr }[({\mathbf 1}\otimes B_i^*)\rho]=\sum_{k=1}^r\sqrt{\lambda_{k}}{\rm tr }[A_k]{\rm tr }[B_i^*B_k]=\sqrt{\lambda_{i}}{\rm tr }[A_i],\\
&&{\rm tr }[(A_i^*\otimes B_i^*)\rho]=\sum_{k=1}^r\sqrt{\lambda_{k}}{\rm tr }[A_i^*A_k]{\rm tr }[B_i^*B_k]=\sqrt{\lambda_{i}}.
\end{eqnarray*}
It then follows that $\sqrt{\lambda_{i}}{\rm tr }[A_i]{\rm tr }[B_i]=1$ for all $i=1,\cdots,r$. But we have 
$$
{\rm tr }[\rho]=\sum_{i=1}^r\sqrt{\lambda_{i}}{\rm tr }[A_i]{\rm tr }[B_i]=1.
$$
This forces $r=1$ and leads to the conclusion $\rho=\sqrt{\lambda_1}A_1\otimes B_1$. 
To complete the proof, we observe that $\rho$ being a mixed state, here only the case 
$$
A_1\in {\cal L}_{\rm s}({\cal H}_1)^{+}\ \hbox{and}\  B_1\in {\cal L}_{\rm s}({\cal H}_2)^{+}
$$
or the case
$$
(-A_1)\in {\cal L}_{\rm s}({\cal H}_1)^{+}\ \hbox{and}\  (-B_1)\in {\cal L}_{\rm s}({\cal H}_2)^{+}
$$
is allowed. Whichever case arises, evidently we can write $\rho=\rho_1\otimes\rho_2$, where $\rho_1$ and $\rho_2$ are mixed states of ${\cal H}_1$ and ${\cal H}_2$ respectively.

Definition 4.5. Let $\rho$ be a generic mixed state of the Hilbert space $\otimes_{k=1}^{n}{\cal H}_k(n\ge 2)$. If
$$
{\rm tr }[(\otimes_{i=1}^{n} P_i)\rho]=\prod\nolimits_{i=1}^{n}{\rm tr }[\overline{P}_i\rho]
$$
for all $P_i\in {\cal P}({\cal H}_i)(i=1,2,\cdots,n)$, we say that $\rho$ satisfies the unentanglement condition of $\otimes_{k=1}^{n}{\cal H}_k$.

Notice that the unentanglement condition in this definition is stronger than in Definition\nobreak{} 3.5. Here, ``for all $P_i\in {\cal P}({\cal H}_i)$" is required instead of ``for all $P_i\in {\cal P}_1({\cal H}_i)$".

Lemma 4.2. Let $\rho$ be a generic mixed state of the Hilbert space $\otimes_{k=1}^{n}{\cal H}_k(n\ge 2)$. If $\rho$ satisfies the unentanglement condition of $\otimes_{k=1}^{n}{\cal H}_k$, then viewed as a mixed state of the Hilbert space ${\cal H}_1\otimes ({\cal H}_2\otimes\cdots\otimes {\cal H}_n)$ it satisfies the following condition:
$$
{\rm tr }[(P\otimes Q)\rho]={\rm tr }[(P\otimes {\mathbf 1})\rho]{\rm tr }[({\mathbf 1}\otimes Q)\rho]
$$
for all $P\in {\cal P}_1({\cal H}_1),Q\in {\cal P}_1(\otimes_{k=2}^{n}{\cal H}_k)$.

Proof. It is easy to verify that the lemma is true when the operator $Q$ is a linear combination of operators of the form $\otimes_{i=2}^{n}P_i,P_i\in {\cal P}_1({\cal H}_i)(i=2,\cdots,n)$, thanks to the legitimateness of taking $P_1={\mathbf 1}$ in the (stronger) unentanglement condition defined in Definition 4.5.

Generally the operator $Q$ takes the form $P_{\Psi},\Psi\in\otimes_{k=2}^{n}{\cal H}_k$. As is shown in the proof of Lemma 3.3, such $Q$ can be approximated by linear combinations of operators of the form just mentioned. In other words, for all $\epsilon>0$ there exists a $Q^{\prime}$, which is a linear combination of operators of the form $\otimes_{i=2}^{n}P_i,P_i\in {\cal P}_1({\cal H}_i)(i=2,\cdots,n)$, such that $\|Q-Q^{\prime}\|<\epsilon$. Now the proof can be completed in the same way as in Lemma 4.1.

Theorem 4.1. Let $\rho$ be a generic mixed state of the Hilbert space $\otimes_{i=1}^{n}{\cal H}_i(n\ge 2)$. Then $\rho$ is a product mixed state if and only if it satisfies the unentanglement condition of $\otimes_{i=1}^{n}{\cal H}_i(n\ge 2)$.

Proof. The necessity is trivial. For the sufficiency we use induction on $n$. The case $n=2$ has been settled in Proposition 4.3. Let $n=k+1$. On account of Lemma 4.2 we can apply Proposition 4.3 to write  $\rho=\rho_1\otimes\rho_2$, where $\rho_1$ is a mixed state of ${\cal H}_1$ and $\rho_2$ a mixed state of $\otimes_{i=2}^{k+1}{\cal H}_i$. $\rho$ satisfying the unentanglement condition of $\otimes_{i=1}^{k+1}{\cal H}_i$, we now have 
$$
{\rm tr }[(\otimes_{i=1}^{k+1} P_i)(\rho_1\otimes\rho_2)]=\prod_{i=1}^{k+1}{\rm tr }[\overline{P}_i(\rho_1\otimes\rho_2)]
$$
for all $P_i\in {\cal P}({\cal H}_i)(i=1,2,\cdots,k+1)$. Taking $P_1={\mathbf 1}$ and using ${\rm tr }[\rho]=1={\rm tr }[\rho_1]$, we arrive at 
$$
{\rm tr }[(\otimes_{i=2}^{k+1} P_i)\rho_2]=\prod_{i=2}^{k+1}{\rm tr }[\overline{P}_i\rho_2]
$$
for all $P_i\in {\cal P}({\cal H}_i)(i=2,\cdots,k+1)$. This is none other than the statement that $\rho_2$ satisfies the unentanglement condition of $\otimes_{i=2}^{k+1}{\cal H}_i$. So the standard induction procedure can be used to complete the proof.

We note that Theorem 4.1 remains true when the unentanglement condition satisfied by $\rho$ is weakened to the following version:  for arbitrary $t\in\{2,\cdots,n\}$ and $1\le i_1<\cdots<i_t\le n$, 
$$
{\rm tr }[(\otimes_{i=1}^{n} P_i)\rho]=\prod_{i=1}^{n}{\rm tr }[\overline{P}_i\rho]
$$
when  $P_i\in {\cal P}_1({\cal H}_i)$ for $i\in\{i_1,\cdots,i_t\}$ and $P_i={\mathbf 1}$ for $i\notin\{i_1,\cdots,i_t\}$. This condition may appear clumsy. Nevertheless the idea behind it should be clear.
\vskip 10pt
\noindent 5. Concluding Remarks
\vskip 5pt
The motivation of this paper comes from physics. So it is desirable to interpret physically the main result of this paper. To this end the following brief discussions are devoted. 

In the conventional accounts of quantum theory [2], a quantum system $S$ is described by a Hilbert space, say ${\cal H}$, a physical state of $S$ is represented by a pure state $\phi$ of ${\cal H}$ or a mixed state $\rho$ of ${\cal H}$, as defined in the previous sections, $\phi$ being identified with $P_{\phi}$ in most cases. A physical state which is not a product state is called an entangled state. As a matter of fact, entanglement is a crucial concept in quantum physics and plays a pivotal role in the field of quantum information [3]. A bounded observable of $S$ is determined by an operator $A$ in ${\cal L}_s({\cal H})$. Let $E$ be the spectrum measure of $A$. Then the probability that a measurement of $A$ in the state $\rho$ gives a result in the Borel set $\Delta$ is supposed to be ${\rm tr }[E(\Delta)\rho]$. Let us denote by $A^{\rho}$ the random variable with the probability distribution defined by ${\rm tr }[E(\Delta)\rho]$.

Now let $S$ be a so called composite system, consisting of the subsystems $S_1,\cdots,S_n$ whose Hilbert spaces are ${\cal H}_1,\cdots,{\cal H}_n$ respectively. Then the Hilbert space ${\cal H}$ of $S$ is ${\cal H}_1\otimes\cdots\otimes{\cal H}_n$ and each bounded observable $A_i\in{\cal L}_s({\cal H}_i)$ of the subsystem $S_i$ is regarded as the observable $\overline{A}_i\in {\cal L}_s({\cal H})$ of the composite system $S$ for all $i=1,\cdots,n$. According to quantum theory, for a physical state $\rho$ of $S$, the probability that the random variable $\overline{A}_i^{\rho}$, on ${\mathbb R}^n$ by definition, takes value in the Borel set $\Delta$ is ${\rm tr }[\overline{E}_i(\Delta)\rho]$, 
and the joint distribution of these random variables is determined in the following way: the probability that $\overline{A}_i^{\rho}$ takes value in the Borel set $\Delta_i\subset{\mathbb R}(i=1,\cdots,n)$ is ${\rm tr }[(E_1(\Delta_1)\otimes\cdots\otimes E_n(\Delta_n))\rho]$. Here $E_i$ is the spectrum measure of $A_i(i=1,\cdots,n)$.

Finally we are prepared to present the physical interpretation of the main result of this paper: if $\rho$ is a generic state of a composite quantum system consisting of $n$ subsystems, then $\rho$ is a product state if and only if the random variables $\overline{A}_i^{\rho}(i=1,\cdots,n)$ are independent for all bounded observable $A_i$ of the $i'$th  subsystem $(i=1,\cdots,n)$. This is doubtlessly rather a natural result, from physical point of view.
\vskip 10pt
\noindent 6. Statements and Declarations
\vskip 5pt
The author has been studying problems related to quantum measurement and quantum entanglement in the recent 3 years and has no competing interests to declare that are relevant to the content of this article. No funding was received for conducting this study. 
\vskip 10pt
\noindent References
\vskip 5pt
\begin{enumerate}
\item Schmidt, E.: Math. Ann. 83(1907).
\item von Neumann, J.: Mathematische Grundlagen der Quantenmechanik, Springer, Berlin, 1932.
\item Nielsen, M. A. and Chuang, I. L.: Quantum Computation and Quantum Information, Cambridge University Press, 2000.
\end{enumerate}

\end{document}